\documentclass[conference,twocolumn,a4paper]{IEEEtran}
\IEEEoverridecommandlockouts
\usepackage{cite}
\usepackage{amsmath,amsfonts,amssymb,amsthm}
\usepackage{algorithmic}
\usepackage{graphicx}
\usepackage{textcomp}
\usepackage{xcolor}
\usepackage[utf8]{inputenc}
\usepackage{commath}
\usepackage{scalerel}
\usepackage{tikz}
\usepackage[]{changes}
\usepackage{caption}
\usepackage{subcaption}
\usepackage{layouts}
\usepackage[draft]{hyperref}

\usepackage{pgfplots}
\DeclareUnicodeCharacter{2212}{−}
\usepgfplotslibrary{groupplots,dateplot}
\usetikzlibrary{patterns,shapes.arrows}
\pgfplotsset{compat=newest}

\def\mystrut{\vphantom{hg}}

\pgfplotsset{
    legend image with text/.style={
        legend image code/.code={%
            \node[anchor=center] at (0.3cm,0cm) {#1};
        }
    },
}

\def\BibTeX{{\rm B\kern-.05em{\sc i\kern-.025em b}\kern-.08em
    T\kern-.1667em\lower.7ex\hbox{E}\kern-.125emX}}

\begin{document}

\title{
Drone delivery: Reliable Cellular UAV Communication Using Multi-Operator Diversity
\thanks{This research is supported by the Research Foundation Flanders (FWO), projects no. S003817N (OmniDrone) and G098020N.}
}


\author{\IEEEauthorblockN{
Achiel Colpaert, 
Micha\"el Raes,
Evgenii Vinogradov, 
and Sofie Pollin 
}
\IEEEauthorblockA{KU Leuven, ESAT - Department of Electrical Engineering, Kasteelpark Arenberg 10, 3001 Heverlee, Belgium\\
E-mail: \{achiel.colpaert, evgenii.vinogradov, sofie.pollin\}@kuleuven.be} michael.raes@student.kuleuven.be}

\maketitle

\begin{abstract}
The market size of Unmanned Aerial Vehicles (UAVs, a.k.a drones) can reach up to 10\% of the global market value. In particular, drone delivery is one of the most attractive applications. The growing number of drones requires appropriate traffic management systems that will rely on cellular networks. However, it has been shown in the literature that these networks cannot provide reliable communication due to low coverage probability and frequent handovers. This article presents a potential solution targeting these problems while requiring no modifications of the existing infrastructure. Namely, equipping the UAV with multiple cellular modems to connect to different providers' networks introduces network diversity resulting in 98\% coverage probability at the flight altitude of 100 meters. In contrast, one network ensures only 80\% coverage. At the same time, the size of the outage zones becomes up to ten times smaller and the frequency of harmful handovers is reduced to zero. The results are obtained with a physical-layer simulator utilizing a real urban 3D environment, cellular network parameters (e.g., site locations, antenna orientation and gains), and specific aerial channel models.
\end{abstract}


\section{Introduction}
 In May 2021, Morgan Stanley released a forecast \cite{MS} stating that by 2050 the total market of Urban Air Mobility (UAM, including drone delivery, air taxi, patrolling drones, to name a few) will reach up to 11\% of the projected global Gross Domestic Product (GDP). The same document claims that the large-scale deployments of drone-based urban delivery are expected to be in place by 2030. 
 
Apart from the mobility, Unmanned Aerial Vehicles and Systems (UAVs, UASs) are expected to play an essential role in future 6G networks where aerial Base Stations (BSs) will provide connectivity to ground users \cite{Tut18, azari2021evolution, UAV_6G}. Shorter-term predictions say that by 2025, the drone services market will be worth a total of USD 63.6 billion \cite{business2021drone} and the UAV fleet, both recreational and commercial, is projected to reach 2 to 3 million by 2023 \cite{faa2019prediction}.

Managing such a large fleet requires the design of UAS Traffic Management (UTM) solutions in order to ensure expected levels of safety, security, operation transparency, and airspace usage efficiency \cite{Dr_Tech}. The UTM framework formulated by International Civil Aviation Organization (ICAO) in \cite{ICAO} includes cellular technology as a critical enabler of large-scale drone deployments. Several competing UTM implementations have been proposed \cite{Bauranov}. All of them rely on the existence of a reliable command and control (C2) link. Indeed, cellular networks are an ideal candidate as they provide ubiquitous coverage and remove the need for the UTM provider to deploy expensive dedicated wireless infrastructure. 3GPP has taken the first steps to introduce aerial vehicles some time ago now \cite{3gpp2018enhanced}, but many questions remain open \cite{Tut18, UAV_6G}.

\subsection{State of the Art: overview of open problems}
\par In an effort to evaluate the performance of UAVs in cellular networks many different types of studies have been performed, from analysis to simulator based, to actual experimental work. Several relevant overviews can be found in \cite{Tut18, azari2021evolution}. In this article, let us provide just a few of the most relevant works.  

An in-depth analysis of the performance of UAVs in a cellular network based on stochastic geometry provided in \cite{azari2019cellular}. Authors of \cite{rodriguez2021air} and \cite{bertold} characterized the UAV channels through dedicated measurement campaigns. All works listed above conclude that the antenna configuration at BSs should be considered when introducing aerial users to a cellular network as the antennas are pointed towards the ground generally. 

Other simulation-based research has proven that current LTE networks do not provide sufficient coverage at altitudes above building height, mainly due to interference problems \cite{colpaert2018aerial} and high handover rates \cite{colpaert2020beamforming}. Several measurement campaigns \cite{gharib2021exhaustive, Fakhreddine2019HandoverCF} showed satisfying results both in terms of coverage and handover rates. However, they were performed in rural areas where the BS density is much lower compared to an urban scenario. 

\textbf{Takeaways:} A high density of base stations results in high signal strengths but simultaneously high interference levels due to line-of-sight conditions. Moreover, the BS antennas are downtilted to optimize the ground coverage. Consequently, the problem of achieving a highly reliable connection with UAVs in an urban environment remains open due to i) limited coverage caused by high interference (growing with increasing flight altitude); ii) frequent handovers. 
\subsection{Multi-Operator Diversity as a Potential Solution}
This article suggests a solution to increase the reliability of UAV communication links while requiring no modifications of the terrestrial cellular networks. Several studies have shown through network level measurements that equipping a UAV with multiple LTE modems to connect to different providers' networks will improve reliability and Quality of Service \cite{amorim2019improve,sae2020reliability,bacco2022airtoground}. Thus, we can assume that introducing network diversity improves the coverage and handover rates experienced by UAVs. Indeed, network infrastructure from different providers is deployed at different base station sites in the city\footnote{Even when the sites are shared (for instance the same mast can be used), the sector antennas are usually pointed differently. This is due to the need to optimize the performance under different underlying network topology.}. This feature reduces the probability of having a bad connection to all BSs at a given location of UAV. 

The price to pay is a slightly higher payload weight since the modern modems are quite compact. The size and weight can be reduced even further if a dedicated multi-connection module is designed. Another factor is an increased power consumption. However, a more stable connection results in less frequent modifications of the flight path \cite{7888557, Sibren19}. Note that power consumption of the communication modems is negligible in comparison with the propulsion energy \cite{7888557}.
\subsection{Contributions}
The main contributions of this work are:
\begin{enumerate}
    \item We designed a realistic simulator taking into account
    \begin{itemize}
        \item real 3D environment including information about i) ground surface, ii) buildings and infrastructure;
        \item real cellular network parameters (e.g., BS locations, sector orientations, used power, etc.) reported by the Belgian operators to the government;
        \item 3D antenna patterns (with adaptable sidelobes);
        \item specific UAV channel models;
        \item users' mobility (UAVs).
    \end{itemize}
    \item We assessed potential effects of multi-operator diversity on i) coverage and ii) handovers.
\end{enumerate}
Note that in this work, we focus on a promising use case of drone delivery. UAV delivery will probably be performed at higher altitudes than other popular operations (e.g., patrolling) due to safety reasons. Though the higher altitude can result in more severe damage in case of malfunction, it gives a better time margin for the safety systems to act. For this reason we use altitudes higher than foreseen in \cite{ICAO}. Of course, the presented results are useful for other UAV applications when only the appropriate altitudes are analyzed.
\section{Model}
\label{sec:model}

\par To model the scenario of drone deliveries, we consider drones flying in an urban environment at a fixed speed. They travel at constant altitudes ranging from just above rooftop level up to a maximum of 300~m above ground level. They fly in a straight line starting at a random point $A$ in the city to another random point $B$ representing the delivery address. The requirements for a network supporting these drones are twofold. On one side, only a low throughput link is necessary to monitor and control the UAV from a remote location. The requirement for this is a minimum Signal-to-Noise-Ratio (SINR). On the other hand, the link needs to be very reliable. Due to safety concerns, one cannot afford to lose connection to the UAV. The amount of handovers (HO) between sectors and radio link failures (RLF) will affect the time that the UAV is not connected to the network.
\par For the location of our study, we choose the city of Leuven as the environment as this is the location of our university, and this allows us to verify our results even further in the future.
\par We will evaluate the performance of the network using several metrics based on the SINR which is calculated at any 3D point $\vec{p}$ in the simulator as follows: \begin{equation}
	\label{eq:sinr_interference_noise}
	SINR (\vec{p}) = \frac{P_r(\vec{p},a)}{\sum^{A}_{i \neq a} P_{r, i}(\vec{p}) + N},
\end{equation}
where $A$ represents the set of all sectors in the environment and $a$ represents the currently assigned sector, $N$ represents the noise power and is calculated as follows $N = BN_0F$, where $B$ is the bandwidth, $N_0$ is the noise density and $F$ is the noise figure. 
\par The metrics considered are the following: the first being coverage probability $P_{cov}$ calculated as the ratio of the area where the Signal-to-Interference-and-Noise-Ratio (SINR) is larger than a threshold (T) based on throughput requirements, $SINR > T$, divided by the total area. The second metric is the size of the largest continuous zone where no coverage exists in square kilometers, called the maximum outage zone $OUT_{max}$. This metric represents the importance of continuous coverage, large regions of outage results in the UAV being disconnected for long periods of time which is detrimental in a reliable network.

\section{Simulations}
\par To simulate the received powers at any 3D point, an improved version of the physical layer simulator developed in \cite{colpaert2018aerial, colpaert2020beamforming} is used. It uses a real 3D environment based on surface scans to create a virtual environment where assets like base stations and users can be deployed in a 3D space. The modified version includes realistic cellular network configurations as we detail in the following subsection.
\subsection{Environment}
\begin{figure}
	\centering
    \includegraphics[scale=0.6]{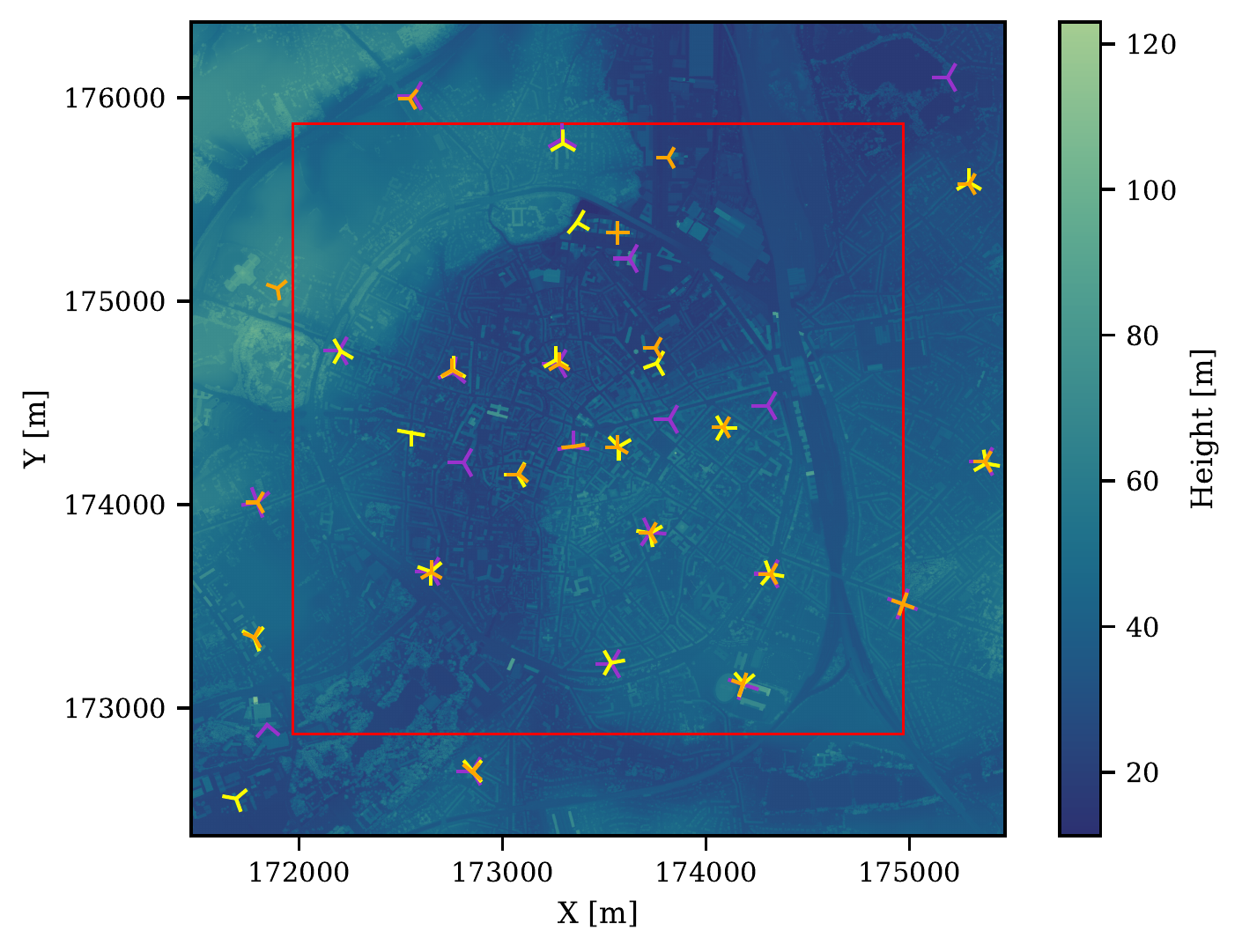}
	\caption{Digital Surface Model with sectors and simulation area indicated in red. The lines represent the sectors and their azimuth rotation and each color represents a different network.}
	\label{fig:leuven_dsm_sectors}
\end{figure}

The simulator requires several parameters based on the real world environment as input. First of all, terrain and surface information is required, as this affects the heights of the antennas and, more importantly, allows us to determine if a sector is in line of sight with a point. For this, the DHMV-dataset, see \cite{colpaert2018aerial}, is used and the considered environment is displayed in Fig.~\ref{fig:leuven_dsm_sectors}.

\par Secondly, information about the location of the different base stations, their sectors, tilts, transmit powers and antennas needs to be known. For this, information from the Belgian Antenna Registry of the BIPT \cite{bipt-ant}, the organisation managing the spectrum in this environment, is combined with the database of \emph{declarations of conformity} issued by operators for each antenna site. These declaration documents contain all information about the hardware deployed at each antenna site and are used as input for the simulator.

\par Thirdly, we used a 3D antenna pattern  corresponding to the antennas used on the sites in the evaluated area \cite{huaweiantenna}. It is important to note that just like the real antenna, this antenna pattern has sidelobe suppression implemented meaning that the actual gain of the sidelobes is further reduces to improve power efficiency by reducing energy emitted towards the sky. 

\par Lastly, the used channel model is defined in \cite{3gpp2018enhanced} specifically for aerial vehicles.

\subsection{Scenarios}
\par Two different scenarios are simulated. One static scenario where all points on the map are evaluated at different altitudes, generating statistics of the area at different heights. Secondly, a mobile scenario is evaluated where a UAV travels a random path over the city populated by static ground users. In this scenario, we can evaluate handover behaviour and Radio-Link-Failure (RLF) occurrences of the network.

\subsubsection{Static scenario}

\par The first scenario's goal is to evaluate the statistics of the chosen city of Leuven and characterize the cellular network at different altitudes above ground level to give us some insight on how well a UAV can utilize the cellular network. Using the info of the environment and the general parameters used in a typical LTE deployment, see Table \ref{tab:general_simparams}, the simulator evaluates the channel model for every 3D point in the simulation area, and as such can calculate the received power from each sector at each 3D point. This happens for three different network operators (Op1, Op2 \& Op3). The base station sites of these operators can coincide. The received power from all sectors can be used to determine cell allocation and SINR levels. SINR levels can then be further used to calculate coverage.

\par This characterization of the environment is performed for different operators. After combining all data, we can calculate the SINR at any point in a multi-operator network, as in \cite{nguyen2017using}:
\begin{equation}
	\label{eq:sinr_max}
	SINR (\vec{p}) = \max\limits_{k}\Bigg(\frac{P_r(\vec{p},a)}{\sum^{A}_{i_k \neq a} P_{r, i_k}(\vec{p}) + N}\Bigg),
\end{equation}
where $k$ is the operator index. These SINR values can then be used to calculate the coverage as described in Section \ref{sec:model}. Using all the datapoints at a specific altitude we can also evaluate the area size of continuous outage zones and we can calculate the largest outage zone, $OUT_{max}$.


\begin{table}[h!]
	\begin{center}
	  \caption{Simulation Parameters}
	  \label{tab:general_simparams}
	  \begin{tabular}{|l|c|} 
		\hline
		\textbf{Parameter} & \textbf{Value}\\
		\hline
		Simulation area, & $9~km^2$\\
		Resolution, & $5~m$\\
		Carrier frequency, $f_c$& $1800~MHz$\\
		Bandwidth, $B$ & $20~MHz$\\
		Outage-threshold, $T$ & $-6~dB$\\
		Noise density, $N_0$ \cite{3gppRadioFrequencyRF2018} & $-174~dBm/Hz$\\
		Noise number, $F$ \cite{3gppRadioFrequencyRF2018} & $9~dB$\\
		\hline

	  \end{tabular}
	\end{center}
  \end{table}
  
\subsubsection{Mobile scenario}

\par Finally a second scenario is simulated to get more insight into the mobility characteristics of a multi-network approach. In \cite{colpaert2020beamforming}, the simulator was extended with the possibility to consider a moving drone that travels over a certain trajectory in three dimensional space and can hence experience changing network circumstances. They also implemented a simplified handover mechanism based on the \verb+A3+-event from the 4G- and 5G standards \cite{3gpp2021NR}.

\par Alongside the handover event, there exists another event that could cause temporary disconnection and latency issues, the
RLF event. When the UE detects problems in the connection it will wait for a specific timer (\verb+T310+ \cite{3gpp2021NR}) to run out and when the problem is not resolved at that time the UE will consider itself in RLF and begin the cell selection procedure again. The handover threshold and the RLF timer are network parameters that the operator can tweak on a per-UE basis. Because we want very low latency for drone applications, the following simulation uses a \verb+T310+ of $200~ms$. For the A3-threshold we take $-2~dB$ as in \cite{3gpp2018enhanced}.

\par In a multi-network context the definitions are slightly different. First, a RLF is defined as an event when the user is not connected to any network. Secondly, a handover is defined as:
\begin{itemize}
	\item One of the used networks has a HO and the other used networks are in RLF at that time;
	\item All used networks experience a handover within 1 second.
\end{itemize}
This period of $1~s$ is self-defined and is a measure for how robust one wants to protect the combined link against near simultaneous handovers. A longer period gives more margin, a shorter period assumes that the handover is correctly resolved when the other network starts a handover. The remainder of the simulation parameters can be found in Table \ref{tab:leuven_ho_stats}, refer to \cite{colpaert2020beamforming} for the full explanation of all the parameters.

\begin{table}[h!]
	\begin{center}
	  \caption{Handover Simulation Parameters}
	  \label{tab:leuven_ho_stats}
	  \begin{tabular}{|l|c|} 
		\hline
		\textbf{Parameter} & \textbf{Value}\\
		\hline
		Heights (AGL) $h$  & $20$~m, $40$~m, $80$~m, $160$~m\\
		Drone velocity $v$  & $15~m/s$\\
		Outage-threshold $T$ & $-6~dB$\\
		A3 - Offset $T_{A3}$ & $2~dB$\\
		RLF-timer \verb+T310+ & $200~ms$\\
		Simulation timestep $\delta_t$ & $100~ms$\\
		Number simulations & $5000$ \\
		\hline

	  \end{tabular}
	\end{center}
  \end{table}

\section{Results}

\par In this section, we look at the results generated by the simulations described above. We evaluate the metrics at different altitudes to get a better insight into the effect of drone height. We also draw the evaluations under different network loads, but we focus on the $100~\%$ network load as this is the worst-case scenario but the most important scenario when looking at the reliability of the communication link.
\par First, we evaluate the coverage situation, followed by a study of the effects of using multiple operators. 
Next, we evaluate the handover results and verify whether using multiple networks results in an improvement in terms of handovers.

\subsection{Coverage}
\begin{figure}[]
    \centering 
	 \resizebox{0.9\linewidth}{!}{
    \input{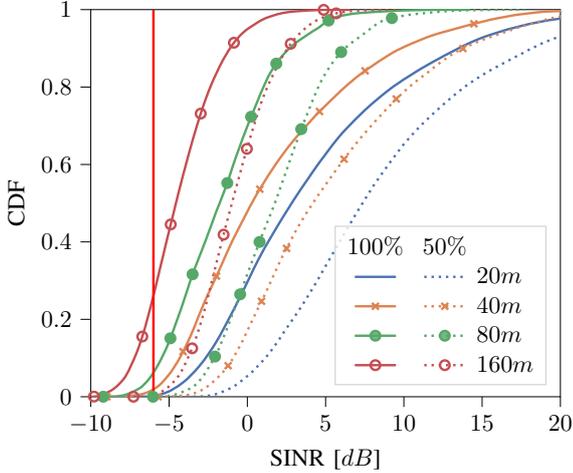}
    }
    \caption{CDF of the SINR of Operator 1 at different heights for a half or full network load. The vertical red line represent the coverage threshold. SINR levels below this threshold are considered as network outage. A network at half load has almost 100\% coverage up to an altitude of $160~m$.}
    \label{fig:leuven_sinr}
\end{figure}
\par The resulting statistics of the SINR for Operator 1 can be seen in Fig. \ref{fig:leuven_sinr} where the CDF of the SINR for different altitudes and loads is shown. To meet the requirements for downlink communications, UAVs must have a minimum SINR of $-6~dB$ \cite{colpaert2018aerial} for a command-and-control link. This is the coverage threshold $T$, below which the drone experiences a SINR that is too low to achieve successful communication. This is indicated by a vertical red line.

\par The figure shows that as the drone flies higher, we can see that the density function shifts to the left, indicating that SINR values drop. One can also see that the curve becomes steeper indicating that the variance keeps decreasing with height: a very wide distribution at $20~m$ becomes much more concentrated at $160~m$.

\par
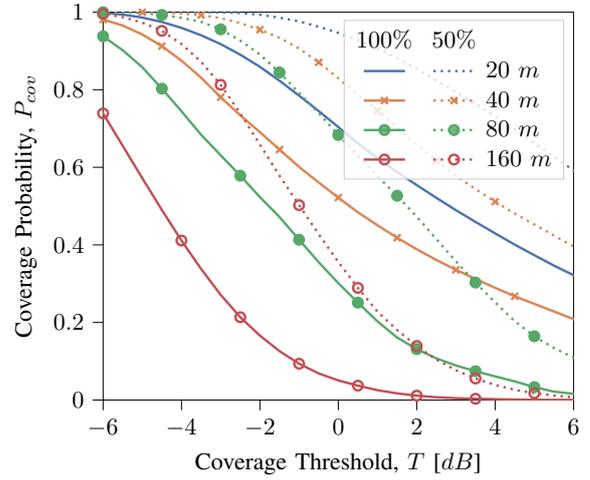
\begin{figure}
	\centering
	 \resizebox{0.9\linewidth}{!}{
\begin{tikzpicture}

\definecolor{color0}{rgb}{0.298039215686275,0.447058823529412,0.690196078431373}
\definecolor{color1}{rgb}{0.866666666666667,0.517647058823529,0.32156862745098}
\definecolor{color2}{rgb}{0.333333333333333,0.658823529411765,0.407843137254902}
\definecolor{color3}{rgb}{0.768627450980392,0.305882352941176,0.32156862745098}

\begin{axis}[
legend cell align={left},
legend style={fill opacity=0.8, draw opacity=1, text opacity=1, draw=white!80!black},
legend columns=2,
legend style={
    font=\mystrut,
    legend cell align=left,
},
tick align=outside,
tick pos=left,
x grid style={white!69.0196078431373!black},
xlabel={Coverage Threshold, $T$ [$dB$]},
xmin=-6, xmax=6,
xtick style={color=black},
y grid style={white!69.0196078431373!black},
ylabel={Coverage Probability, $P_{cov}$},
ymin=0, ymax=1,
ytick style={color=black}
]
\addlegendimage{legend image with text=$100\%$}
\addlegendentry{}
\addlegendimage{legend image with text=$50\%$}
\addlegendentry{}
\addplot [line width=1.0pt, color0]
table {%
-7 0.999977231025696
-6.5 0.999641180038452
-6 0.998011827468872
-5.5 0.993778944015503
-5 0.986085176467896
-4.5 0.974705457687378
-4 0.959611415863037
-3.5 0.940409660339355
-3 0.917675971984863
-2.5 0.890339136123657
-2 0.859145402908325
-1.5 0.823200345039368
-1 0.784802794456482
-0.5 0.744963884353638
0 0.703732013702393
0.5 0.661916375160217
1 0.623131394386292
1.5 0.587850093841553
2 0.554275035858154
2.5 0.521309494972229
3 0.489463448524475
3.5 0.459146976470947
4 0.429608345031738
4.5 0.400633573532104
5 0.373367786407471
5.5 0.346626281738281
6 0.321901440620422
6.5 0.300284266471863
};
\addlegendentry{}
\addplot [line width=1.0pt, color0, dotted]
table {%
-7 1
-3.5 0.999866127967834
-3 0.999324917793274
-2.5 0.997615814208984
-2 0.993337392807007
-1.5 0.985663652420044
-1 0.976215124130249
-0.5 0.962559461593628
0 0.946608066558838
0.5 0.92812705039978
1 0.906381726264954
1.5 0.882081270217896
2 0.85400390625
2.5 0.823311448097229
3 0.794575929641724
3.5 0.763923406600952
4 0.731695652008057
4.5 0.697926878929138
5 0.662343621253967
6 0.592154145240784
6.5 0.557505130767822
};
\addlegendentry{$20~m$}
\addplot [line width=1.0pt, color1, mark=x, mark repeat={3}, mark options={solid}]
table {%
-7 0.995241403579712
-6.5 0.990424513816833
-6 0.980918645858765
-5.5 0.965644478797913
-5 0.942358016967773
-4.5 0.912482976913452
-4 0.874948501586914
-3.5 0.831153869628906
-3 0.780678749084473
-2.5 0.733933329582214
-2 0.690728187561035
-1.5 0.645623564720154
-1 0.601058483123779
-0.5 0.560723543167114
0 0.521915435791016
0.5 0.483930587768555
1 0.450548410415649
1.5 0.418470740318298
2 0.38869297504425
2.5 0.361095666885376
3 0.335737228393555
3.5 0.312261581420898
4 0.2893226146698
4.5 0.267479538917542
6 0.208658218383789
6.5 0.190708756446838
};
\addlegendentry{}
\addplot [line width=1.0pt, color1, dotted, mark=x, mark repeat={3}, mark options={solid}]
table {%
-7 1
-5 0.999666213989258
-4.5 0.998450875282288
-4 0.996201038360596
-3.5 0.992630004882812
-3 0.984634160995483
-2.5 0.971801996231079
-2 0.954736828804016
-1.5 0.933683633804321
-1 0.906353235244751
-0.5 0.870607137680054
0 0.829646468162537
0.5 0.787314653396606
1 0.746014595031738
1.5 0.702292203903198
2 0.658163785934448
2.5 0.61738383769989
3 0.578600764274597
3.5 0.543644547462463
4 0.511425018310547
4.5 0.480882406234741
6 0.395854473114014
6.5 0.369149923324585
};
\addlegendentry{$40~m$}
\addplot [line width=1.0pt, color2, mark=*, mark repeat={3}, mark options={solid}]
table {%
-7 0.980716705322266
-6.5 0.964327812194824
-6 0.937661170959473
-5.5 0.901883363723755
-5 0.858041644096375
-4.5 0.802763938903809
-4 0.744447231292725
-3.5 0.683861136436462
-2.5 0.578400015830994
-2 0.523316621780396
-1.5 0.471825003623962
-1 0.413255572319031
-0.5 0.355283260345459
0 0.301361083984375
0.5 0.25106942653656
1 0.202663898468018
1.5 0.161088943481445
2 0.131269454956055
2.5 0.107397198677063
3 0.0889694690704346
3.5 0.0742361545562744
4 0.0608916282653809
4.5 0.0480166673660278
5 0.033227801322937
5.5 0.021588921546936
6 0.0159167051315308
6.5 0.0122444629669189
};
\addlegendentry{}
\addplot [line width=1.0pt, color2, dotted, mark=*, mark repeat={3}, mark options={solid}]
table {%
-7 1
-6 0.999997138977051
-5.5 0.999533414840698
-5 0.997600078582764
-4.5 0.99286675453186
-4 0.9852694272995
-3.5 0.97504448890686
-3 0.955780506134033
-2.5 0.926708340644836
-2 0.890936136245728
-1.5 0.844488859176636
-1 0.791713953018188
-0.5 0.739855527877808
0 0.683194398880005
0.5 0.629683256149292
1 0.576952695846558
1.5 0.526600003242493
2 0.47278892993927
3 0.357519388198853
3.5 0.302783370018005
4 0.251888871192932
4.5 0.203147172927856
5 0.164294481277466
5.5 0.134108304977417
6 0.109505534172058
6.5 0.0909055471420288
};
\addlegendentry{$80~m$}
\addplot [line width=1.0pt, color3, mark=o, mark repeat={3}, mark options={solid}]
table {%
-7 0.87862491607666
-6.5 0.816091656684875
-6 0.73895001411438
-5 0.572083353996277
-4.5 0.489966630935669
-4 0.411166667938232
-3.5 0.337991714477539
-3 0.271308302879333
-2.5 0.213508367538452
-2 0.166169404983521
-1.5 0.126036167144775
-1 0.093447208404541
-0.5 0.0684916973114014
0 0.0505222082138062
0.5 0.0370138883590698
1 0.0253027677536011
1.5 0.0170778036117554
2 0.0112360715866089
2.5 0.00719165802001953
3 0.00472497940063477
3.5 0.00319719314575195
4.5 0.00118339061737061
5 0.000552773475646973
6.5 2.50339508056641e-05
};
\addlegendentry{}
\addplot [line width=1.0pt, color3, dotted, mark=o, mark repeat={3}, mark options={solid}]
table {%
-7 0.999927759170532
-6.5 0.999477863311768
-6 0.995466709136963
-5.5 0.987174987792969
-5 0.973161101341248
-4.5 0.951763868331909
-4 0.917977809906006
-3.5 0.8719722032547
-3 0.812672138214111
-2.5 0.739083290100098
-1.5 0.578566670417786
-1 0.502369403839111
-0.5 0.427655577659607
0 0.355913877487183
0.5 0.289049983024597
1 0.227694511413574
1.5 0.180425047874451
2 0.139313936233521
2.5 0.104269504547119
3 0.0766305923461914
3.5 0.0554111003875732
4 0.0388027429580688
4.5 0.0263082981109619
5 0.0175639390945435
5.5 0.0116583108901978
6 0.00716948509216309
6.5 0.00450277328491211
};
\addlegendentry{$160~m$}
\end{axis}

\end{tikzpicture}
	}
	\caption{Impact of the outage threshold, $T$, on the coverage probability, $P_{cov}$ at different altitudes under a $50\%$ or a $100\%$ network load.}
	\label{fig:leuven_cov_vs_threshold}
\end{figure}

 The effect of choosing a different outage threshold can be seen in Fig.~\ref{fig:leuven_cov_vs_threshold}, using the suggested threshold of $-6~dB$ results in a coverage probability of $0.74$ at an altitude of $160~m$. If a larger threshold is chosen the coverage probability drops drastically with $P_{cov}$ reaching zero at a threshold of $2~dB$ for an altitude of $160~m$. These results indicate the bad aerial coverage situation.

\par
\begin{figure}[]
    \centering 
    
	\resizebox{0.9\linewidth}{!}{
    \begin{subfigure}{\linewidth}
\begin{tikzpicture}

\definecolor{color0}{rgb}{0.298039215686275,0.447058823529412,0.690196078431373}
\definecolor{color1}{rgb}{0.866666666666667,0.517647058823529,0.32156862745098}
\definecolor{color2}{rgb}{0.333333333333333,0.658823529411765,0.407843137254902}

\begin{axis}[
legend cell align={left},
legend style={fill opacity=0.8, draw opacity=1, text opacity=1, at={(0.03,0.03)}, anchor=south west, draw=white!80!black},
tick align=outside,
tick pos=left,
x grid style={white!69.0196078431373!black},
xlabel={Height [$m$]},
xmin=0, xmax=300,
xtick style={color=black},
y grid style={white!69.0196078431373!black},
ylabel={Coverage Probability, $P_{cov}$},
ymin=0.5, ymax=1,
ytick style={color=black}
]
\addplot [line width=1.0pt, color0]
table {%
2 0.999848365783691
6 0.999863147735596
10 0.999586582183838
12 0.997876405715942
14 0.996934294700623
16 0.997975945472717
18 0.998380422592163
20 0.998011827468872
22 0.99799108505249
24 0.998309373855591
26 0.998451471328735
28 0.998361229896545
30 0.997753143310547
32 0.996879100799561
34 0.995389938354492
36 0.992586612701416
38 0.988445997238159
40 0.980918645858765
42 0.970294713973999
44 0.968319892883301
46 0.968779444694519
48 0.969600439071655
50 0.96755838394165
52 0.962995409965515
54 0.957619667053223
56 0.949039816856384
58 0.944198250770569
60 0.944767475128174
62 0.944751024246216
64 0.945707440376282
68 0.949952244758606
70 0.948944091796875
72 0.948430299758911
74 0.948441386222839
76 0.947446942329407
78 0.943852782249451
80 0.937661170959473
86 0.924355506896973
88 0.917308330535889
90 0.909850001335144
92 0.902824997901917
96 0.887647151947021
98 0.881272196769714
100 0.873461127281189
102 0.866491675376892
104 0.860411167144775
106 0.856349945068359
108 0.853419423103333
110 0.851569414138794
114 0.8429194688797
116 0.837774991989136
120 0.827866673469543
122 0.823722243309021
124 0.818630576133728
126 0.81196117401123
128 0.803672194480896
130 0.794977784156799
132 0.785152792930603
136 0.772136092185974
138 0.766466617584229
140 0.762786149978638
142 0.759319424629211
146 0.752658367156982
148 0.749902725219727
150 0.746427774429321
152 0.743958353996277
154 0.741883277893066
156 0.740716695785522
160 0.73895001411438
162 0.739269495010376
164 0.73895001411438
168 0.73913049697876
170 0.739536046981812
172 0.739613890647888
174 0.740447282791138
176 0.741761088371277
178 0.743222236633301
180 0.74643611907959
182 0.747616648674011
184 0.747986078262329
188 0.750149965286255
190 0.749655485153198
192 0.74691104888916
194 0.74293053150177
196 0.740299940109253
202 0.734263896942139
204 0.733522176742554
206 0.731177806854248
208 0.727561116218567
210 0.725158333778381
212 0.725230574607849
214 0.725499987602234
216 0.726288914680481
220 0.727002859115601
222 0.726808309555054
224 0.726188898086548
226 0.726555585861206
228 0.72670841217041
230 0.72599720954895
234 0.722130537033081
236 0.719680547714233
238 0.715938806533813
242 0.707886099815369
244 0.704238891601562
250 0.698400020599365
252 0.68060827255249
254 0.674375057220459
256 0.671277761459351
258 0.670102834701538
260 0.670919418334961
262 0.670483350753784
264 0.669125080108643
266 0.667191743850708
272 0.660813808441162
274 0.657369375228882
276 0.653322219848633
278 0.649625062942505
280 0.648708343505859
284 0.647186040878296
286 0.645966649055481
288 0.645822286605835
290 0.644638895988464
294 0.639869451522827
296 0.636311054229736
298 0.632977724075317
300 0.630294442176819
};
\addlegendentry{op1}
\addplot [line width=1.0pt, color1, dashed]
table {%
2 1
38 0.9998859167099
40 0.999644041061401
42 0.999160170555115
44 0.998479127883911
46 0.996949911117554
48 0.997620344161987
50 0.997759580612183
52 0.99660062789917
54 0.994436740875244
56 0.990637063980103
58 0.988345623016357
62 0.984271764755249
64 0.982505321502686
66 0.981910943984985
68 0.981872081756592
72 0.981274843215942
76 0.977877616882324
78 0.97480833530426
80 0.971061110496521
82 0.968269467353821
84 0.966441631317139
86 0.965886116027832
88 0.963663816452026
90 0.960555553436279
92 0.958969473838806
94 0.957572221755981
96 0.957158327102661
98 0.958047151565552
100 0.957294464111328
102 0.954622268676758
104 0.950950026512146
108 0.946836113929749
110 0.947644472122192
112 0.947197198867798
114 0.946005582809448
116 0.944383382797241
118 0.943258285522461
120 0.941961050033569
122 0.941711187362671
124 0.942872285842896
126 0.942744493484497
128 0.940805554389954
130 0.938127756118774
132 0.937769412994385
134 0.937936067581177
136 0.93785548210144
138 0.938063859939575
140 0.937644481658936
142 0.936597228050232
144 0.933880567550659
146 0.9305499792099
150 0.923141717910767
152 0.920697212219238
154 0.919813871383667
156 0.917949914932251
158 0.915502786636353
160 0.911922216415405
162 0.909344434738159
164 0.907291650772095
166 0.904966592788696
170 0.901400089263916
172 0.900008320808411
174 0.898961067199707
176 0.899344444274902
178 0.90084445476532
180 0.901822209358215
188 0.906975030899048
190 0.90785551071167
192 0.907666683197021
194 0.906655550003052
198 0.9032222032547
200 0.902119398117065
202 0.901274919509888
204 0.901058316230774
208 0.901294469833374
210 0.90202784538269
214 0.904716730117798
216 0.904961109161377
218 0.904047250747681
220 0.903666734695435
222 0.903508305549622
224 0.903741598129272
226 0.904947280883789
228 0.905719518661499
230 0.906158328056335
232 0.907161116600037
234 0.907327771186829
236 0.906166672706604
238 0.905430555343628
240 0.905769467353821
242 0.904830574989319
244 0.903172254562378
246 0.902775049209595
248 0.902588844299316
250 0.902602791786194
252 0.899769425392151
254 0.89989161491394
256 0.900877714157104
258 0.902394533157349
260 0.902619361877441
262 0.901744365692139
264 0.900150060653687
266 0.898725032806396
268 0.897050023078918
270 0.894941687583923
272 0.893199920654297
276 0.888044357299805
278 0.885841608047485
280 0.885819435119629
282 0.886797189712524
284 0.887088894844055
286 0.886224985122681
288 0.885591745376587
290 0.884366750717163
292 0.882180571556091
294 0.878775000572205
296 0.874808311462402
298 0.871230602264404
300 0.868858337402344
};
\addlegendentry{op1 \& op2}
\addplot [line width=1.0pt, color2, dotted]
table {%
2 1
44 0.999908208847046
46 0.999510645866394
52 0.999727606773376
54 0.999441385269165
56 0.998546957969666
58 0.998422026634216
64 0.999366641044617
66 0.999316692352295
70 0.998363852500916
72 0.997463941574097
74 0.997174978256226
76 0.996350049972534
78 0.995947241783142
80 0.995208263397217
82 0.99382221698761
84 0.99192214012146
86 0.989163875579834
88 0.987197160720825
90 0.986313819885254
92 0.986816644668579
94 0.986897230148315
96 0.986791610717773
98 0.98645555973053
100 0.985330581665039
102 0.983252763748169
104 0.980561137199402
106 0.979255557060242
108 0.97771942615509
110 0.977452754974365
112 0.976730585098267
114 0.976150035858154
116 0.976000070571899
120 0.978241682052612
122 0.978916645050049
124 0.979305505752563
126 0.979052782058716
128 0.978124976158142
130 0.977402806282043
132 0.975908279418945
134 0.974711179733276
136 0.972641706466675
142 0.97088885307312
144 0.970669507980347
146 0.970013856887817
148 0.969216585159302
154 0.969966650009155
156 0.969908356666565
158 0.969366669654846
160 0.967777729034424
164 0.964149951934814
166 0.962991714477539
168 0.963022232055664
170 0.962683320045471
174 0.961266756057739
176 0.961277723312378
178 0.961791753768921
180 0.962144374847412
182 0.961841583251953
184 0.96197497844696
190 0.964908361434937
194 0.968341588973999
196 0.969247221946716
198 0.969708323478699
200 0.969991683959961
202 0.969783306121826
206 0.969963908195496
208 0.969758272171021
210 0.969408273696899
212 0.969261169433594
216 0.968286037445068
218 0.96797776222229
220 0.96839165687561
222 0.968199968338013
224 0.967472195625305
226 0.9676833152771
228 0.968430519104004
230 0.96940279006958
232 0.970844507217407
234 0.971605539321899
236 0.971763849258423
238 0.971499919891357
242 0.969308376312256
244 0.967716693878174
246 0.966899991035461
248 0.967216730117798
250 0.967747211456299
252 0.963325023651123
254 0.96358335018158
258 0.964577794075012
262 0.963738918304443
264 0.962877750396729
268 0.960522174835205
272 0.95746111869812
274 0.955688953399658
276 0.953202724456787
278 0.950955629348755
280 0.949944496154785
282 0.950205564498901
284 0.949963808059692
286 0.948572158813477
288 0.947780609130859
290 0.947786092758179
292 0.947041749954224
300 0.941291689872742
};
\addlegendentry{op1 \& op2 \& op3}
\end{axis}

\end{tikzpicture}
        \caption{Coverage probability $P_{cov}$}
        \label{fig:leuven_multi_cov}
    \end{subfigure}} 
    \resizebox{0.9\linewidth}{!}{
    \begin{subfigure} {\linewidth}
\begin{tikzpicture}

\definecolor{color0}{rgb}{0.298039215686275,0.447058823529412,0.690196078431373}
\definecolor{color1}{rgb}{0.866666666666667,0.517647058823529,0.32156862745098}
\definecolor{color2}{rgb}{0.333333333333333,0.658823529411765,0.407843137254902}

\begin{axis}[
legend cell align={left},
legend style={fill opacity=0.8, draw opacity=1, text opacity=1, at={(0.03,0.97)}, anchor=north west, draw=white!80!black},
tick align=outside,
tick pos=left,
x grid style={white!69.0196078431373!black},
xlabel={Height [$m$]},
xmin=0, xmax=300,
xtick style={color=black},
y grid style={white!69.0196078431373!black},
ylabel={Max Outage Zone, $OUT_{max}$ [$km^2$]},
ymin=0.0, ymax=3.0,
yticklabel style={%
                 /pgf/number format/.cd,
                     fixed,
                     fixed zerofill,
                     precision=1,},
ytick style={color=black}
]
\addplot [line width=1.0pt, color0]
table {%
2 0.000399947166442871
10 0.000475049018859863
14 0.00325000286102295
24 0.0051499605178833
28 0.00715005397796631
30 0.0100250244140625
32 0.00747501850128174
34 0.00817501544952393
36 0.0168999433517456
38 0.0111500024795532
40 0.029574990272522
42 0.0831999778747559
44 0.0942749977111816
46 0.113499999046326
48 0.082550048828125
50 0.0698250532150269
52 0.0643750429153442
54 0.0882749557495117
56 0.190875053405762
58 0.107025027275085
60 0.113700032234192
62 0.116775035858154
64 0.12279999256134
66 0.131799936294556
68 0.134925007820129
70 0.143025040626526
72 0.146474957466125
74 0.1267249584198
76 0.129624962806702
78 0.149875044822693
80 0.128975033760071
82 0.122974991798401
84 0.144950032234192
86 0.15910005569458
88 0.150349974632263
90 0.233449935913086
92 0.271224975585938
94 0.290349960327148
96 0.427024960517883
98 0.425824999809265
100 0.434550046920776
102 0.516674995422363
104 0.568750023841858
106 0.806249976158142
108 0.815024971961975
110 0.515275001525879
112 0.46922492980957
114 0.48169994354248
116 0.492924928665161
118 0.502900004386902
120 0.586225032806396
122 0.506700038909912
124 0.68417501449585
126 0.788349986076355
128 0.523400068283081
130 0.761950016021729
132 0.84042501449585
134 0.885050058364868
136 1.07057499885559
138 1.11240005493164
140 1.14954996109009
142 1.04050004482269
144 1.07247495651245
146 1.08712494373322
148 1.10047495365143
150 1.13390004634857
152 1.16065001487732
154 1.12779998779297
156 1.1487250328064
158 1.11072504520416
160 1.09702503681183
162 1.04929995536804
164 1.0393500328064
166 1.03957498073578
168 1.02377498149872
170 1.01724994182587
172 1.80522501468658
174 1.90369999408722
176 1.00335001945496
178 0.944875001907349
180 0.931299924850464
182 0.926049947738647
184 1.22510004043579
186 1.21607494354248
188 1.23395001888275
190 1.21157503128052
192 0.892324924468994
194 0.737800002098083
196 0.727450013160706
198 0.725775003433228
200 0.726925015449524
202 0.970900058746338
204 1.7488249540329
206 1.75042498111725
208 1.73682498931885
210 1.83984994888306
212 2.02815008163452
214 0.994849920272827
216 0.97962498664856
218 0.978349924087524
220 0.970799922943115
222 1.66279995441437
224 1.65400004386902
226 0.947624921798706
228 0.801249980926514
230 0.772799968719482
232 1.56105005741119
234 1.6417750120163
236 1.44945001602173
238 1.6432249546051
240 1.52620005607605
242 1.86380004882812
244 1.89744997024536
246 1.78497505187988
248 1.41770005226135
250 1.41592502593994
252 1.70739996433258
254 1.61942505836487
256 1.84957504272461
258 1.87742495536804
260 2.30165004730225
262 2.35500001907349
264 2.45152497291565
266 2.2728750705719
268 2.31837511062622
270 2.35004997253418
272 2.42482495307922
274 2.46997499465942
276 2.50399994850159
278 2.54372501373291
280 2.55987501144409
282 2.81865000724792
284 2.84369993209839
286 2.72392511367798
288 2.6588249206543
290 2.61905002593994
292 2.47822499275208
294 2.62730002403259
296 1.97440004348755
298 2.02522492408752
300 2.45622491836548
};
\addlegendentry{op1}
\addplot [line width=1.0pt, color1, dashed]
table {%
2 0
38 0.000249981880187988
42 0.00412499904632568
44 0.0118000507354736
46 0.0225249528884888
48 0.010949969291687
50 0.0069500207901001
52 0.00870001316070557
54 0.0142999887466431
56 0.0389000177383423
58 0.041700005531311
60 0.0474499464035034
62 0.0500999689102173
64 0.0505249500274658
66 0.0544500350952148
68 0.0488749742507935
70 0.0577499866485596
72 0.0522500276565552
74 0.0284249782562256
76 0.0423500537872314
78 0.0455000400543213
80 0.0453250408172607
82 0.0537500381469727
84 0.100350022315979
86 0.0754749774932861
88 0.070525050163269
90 0.0956250429153442
92 0.103350043296814
94 0.0576000213623047
96 0.0786750316619873
98 0.0601500272750854
100 0.0484499931335449
102 0.0546499490737915
104 0.0735499858856201
106 0.151574969291687
108 0.162199974060059
114 0.182500004768372
116 0.189725041389465
118 0.190999984741211
120 0.187124967575073
122 0.191475033760071
124 0.191550016403198
126 0.187000036239624
128 0.19159996509552
132 0.196500062942505
134 0.203675031661987
136 0.212875008583069
138 0.218950033187866
140 0.221349954605103
142 0.221475005149841
144 0.229550004005432
146 0.230149984359741
148 0.297950029373169
150 0.308099985122681
152 0.322975039482117
154 0.326300024986267
156 0.323300004005432
158 0.317474961280823
160 0.321925044059753
162 0.321449995040894
166 0.336199998855591
168 0.365625023841858
170 0.370924949645996
172 0.38319993019104
174 0.212700009346008
176 0.22409999370575
178 0.231649994850159
180 0.240824937820435
182 0.236474990844727
184 0.22807502746582
188 0.208400011062622
190 0.201375007629395
192 0.205525040626526
194 0.206699967384338
196 0.206650018692017
202 0.223250031471252
204 0.220350027084351
214 0.195425033569336
216 0.192199945449829
218 0.192324995994568
220 0.187549948692322
222 0.177600026130676
226 0.159700036048889
228 0.153674960136414
230 0.145825028419495
232 0.136775016784668
234 0.128900051116943
238 0.117050051689148
242 0.119349956512451
244 0.124974966049194
246 0.126374959945679
250 0.1232750415802
252 0.105725049972534
254 0.12547504901886
256 0.126924991607666
258 0.125125050544739
260 0.121099948883057
262 0.129449963569641
264 0.139925003051758
266 0.120825052261353
268 0.123924970626831
270 0.121175050735474
272 0.187875032424927
274 0.198750019073486
276 0.263224959373474
278 0.274975061416626
280 0.280125021934509
282 0.291000008583069
284 0.28172492980957
286 0.309100031852722
288 0.205450057983398
290 0.311500072479248
292 0.334850072860718
294 0.34404993057251
296 0.349375009536743
298 0.342700004577637
300 0.365625023841858
};
\addlegendentry{op1 \& op2}
\addplot [line width=1.0pt, color2, dotted]
table {%
2 0
44 0.000300049781799316
46 0.00337505340576172
48 0.00139999389648438
54 0.00160002708435059
56 0.00429999828338623
60 0.00594997406005859
62 0.00475001335144043
68 0.00460004806518555
70 0.00670003890991211
72 0.0074000358581543
74 0.00600004196166992
76 0.00765001773834229
80 0.00697505474090576
82 0.0125000476837158
84 0.016124963760376
86 0.0376750230789185
88 0.041949987411499
90 0.0401999950408936
92 0.026824951171875
94 0.0232750177383423
98 0.027649998664856
102 0.0505750179290771
104 0.069350004196167
106 0.114524960517883
110 0.125975012779236
114 0.14014995098114
116 0.14282500743866
118 0.117849946022034
120 0.103950023651123
122 0.10004997253418
124 0.0983250141143799
126 0.0949000120162964
128 0.099174976348877
130 0.100374937057495
134 0.112174987792969
136 0.150449991226196
138 0.157500028610229
140 0.160699963569641
142 0.160074949264526
144 0.144899964332581
146 0.142374992370605
152 0.138450026512146
154 0.145300030708313
156 0.137249946594238
158 0.109549999237061
160 0.102100014686584
162 0.145925045013428
164 0.177049994468689
166 0.18197500705719
168 0.20930004119873
170 0.136500000953674
172 0.144799947738647
174 0.150524973869324
176 0.15137505531311
182 0.134799957275391
184 0.12185001373291
186 0.118399977684021
188 0.117624998092651
192 0.109774947166443
194 0.104974985122681
196 0.0935750007629395
200 0.0839999914169312
202 0.0775500535964966
206 0.0752749443054199
208 0.0729000568389893
210 0.073449969291687
212 0.0726000070571899
214 0.0830999612808228
216 0.10004997253418
218 0.104650020599365
220 0.105649948120117
226 0.0960500240325928
228 0.0908249616622925
232 0.0738500356674194
234 0.0568000078201294
236 0.0500750541687012
238 0.0462249517440796
242 0.0514500141143799
244 0.0574500560760498
246 0.0604749917984009
248 0.0605499744415283
252 0.0557750463485718
254 0.0586249828338623
258 0.0665249824523926
260 0.0679500102996826
262 0.0806000232696533
264 0.0822750329971313
266 0.0876250267028809
268 0.0916750431060791
270 0.0926250219345093
272 0.0904999971389771
276 0.0943249464035034
278 0.0944750308990479
280 0.100275039672852
282 0.102100014686584
284 0.0996999740600586
286 0.102100014686584
288 0.0688749551773071
290 0.0634000301361084
292 0.0542750358581543
294 0.0419249534606934
296 0.0398499965667725
298 0.0291750431060791
300 0.0499500036239624
};
\addlegendentry{op1 \& op2 \& op3}
\end{axis}

\end{tikzpicture}
        \caption{Maximum outage zone $OUT_{max}$}
        \label{fig:leuven_multi_outmax}
    \end{subfigure} 
    }
    \caption{Impact of multi-operator diversity on the network performance. Using multiple cellular networks enables UAV connectivity at higher altitudes.}
\end{figure}
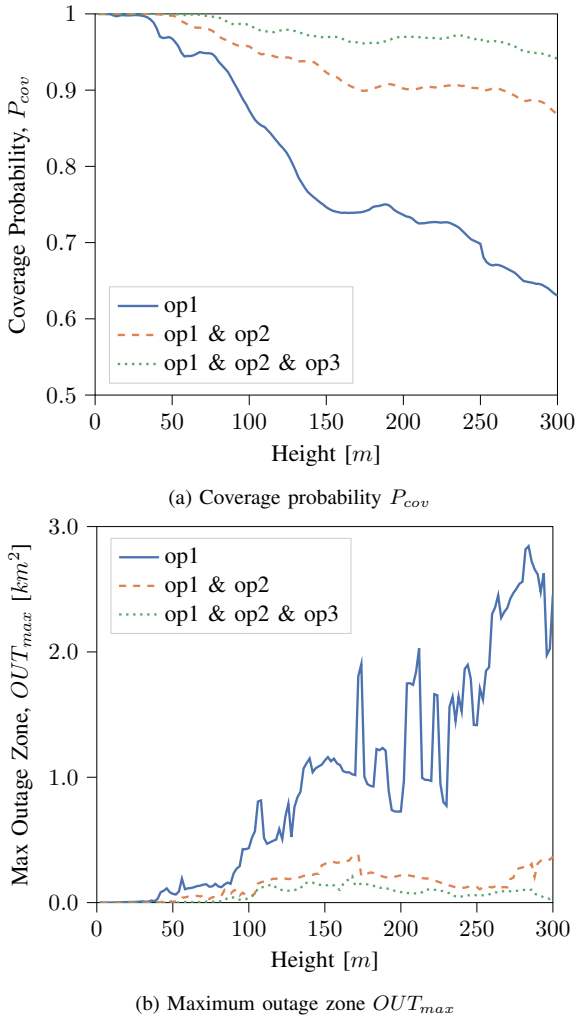

 In an effort to solve this coverage problem, we consider a connection with multiple networks at the same time. The SINR values in this network are calculated using \eqref{eq:sinr_max}.
The coverage probability as a function of the drone altitude can be seen in Fig. \ref{fig:leuven_multi_cov}. When using one network operator, the probability stays one up until around $20~m$ height as the network is designed this way. However, we can see the coverage drops as soon as the UAV flies above rooftop height. In a multi-operator network we can see that the 100\% coverage reaches a much higher altitude. At an altitude of $100~m$ the coverage probability in a multi-operator network is still 0.99. This altitude is much more likely to be used by a drone network. Above this altitude the coverage drops again, albeit less than with one operator. The increase in coverage probability from using two compared to using three operators is less but can be considered for ultra-reliable systems. 

\par Fig. \ref{fig:leuven_multi_outmax} shows the $OUT_{max}$ as function of the altitude. Using a single operator results in relatively large outage zones, especially at altitudes above $100~m$. In a multi-operator scenario the size of the outage zones stays relatively small even at very large altitudes. A single operator network experiences much larger continuous outage zones than a two or three operator network. This can be explained by the fact that many of the antenna sites are different for the selection operators and even if they utilize the same antenna site they often deploy their sectors with different infrastructure, resulting in different azimuths, tilts, etc.

\subsection{Handovers}
\par
\begin{figure}
	\centering
    \begin{subfigure}[b]{0.49\linewidth}
        \centering
        \includegraphics[width=\textwidth]{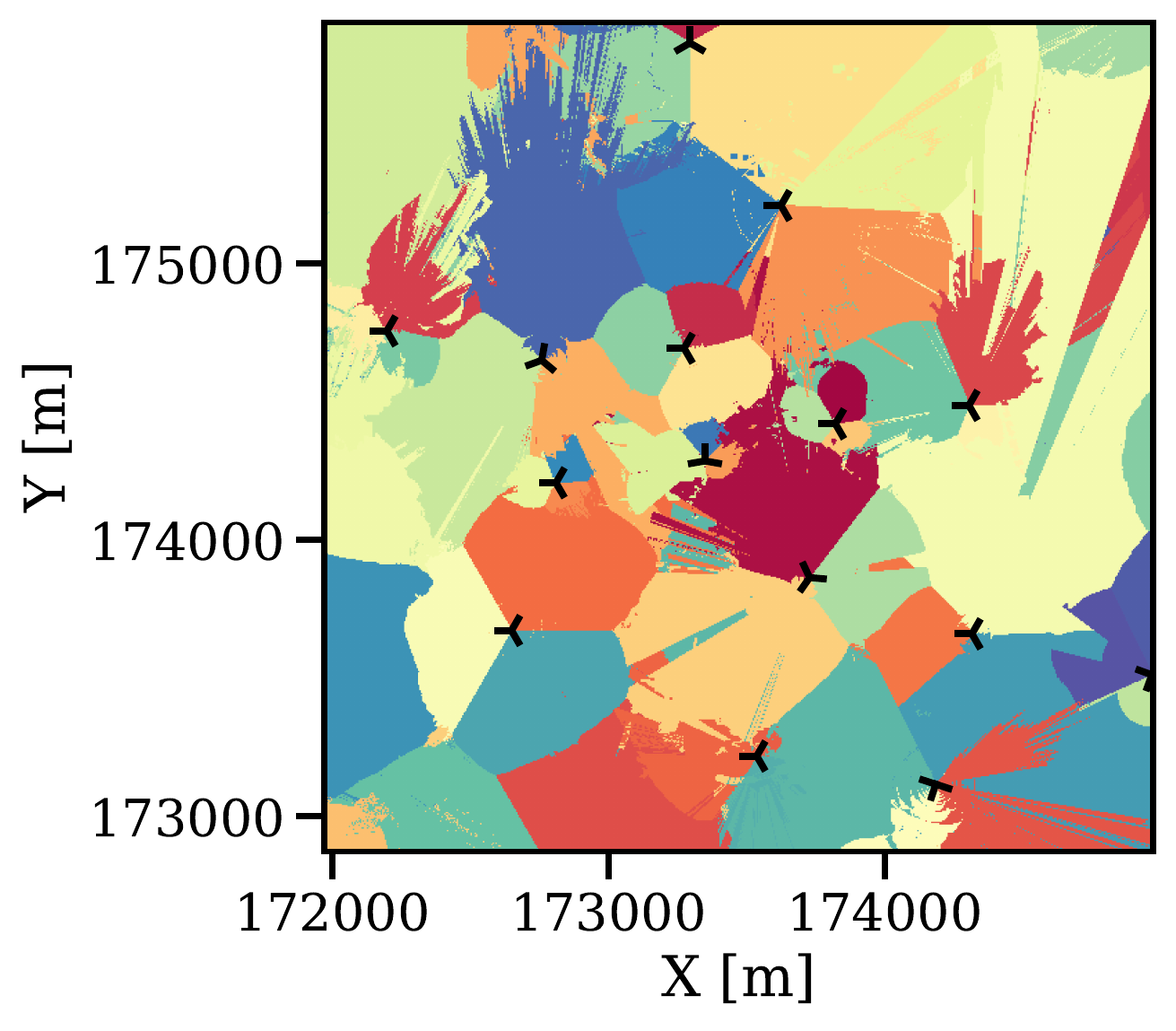}
        \caption{Operator 1, $20~m$}
    \end{subfigure}
    \hfill
    \begin{subfigure}[b]{0.49\linewidth}
        \centering
        \includegraphics[width=\textwidth]{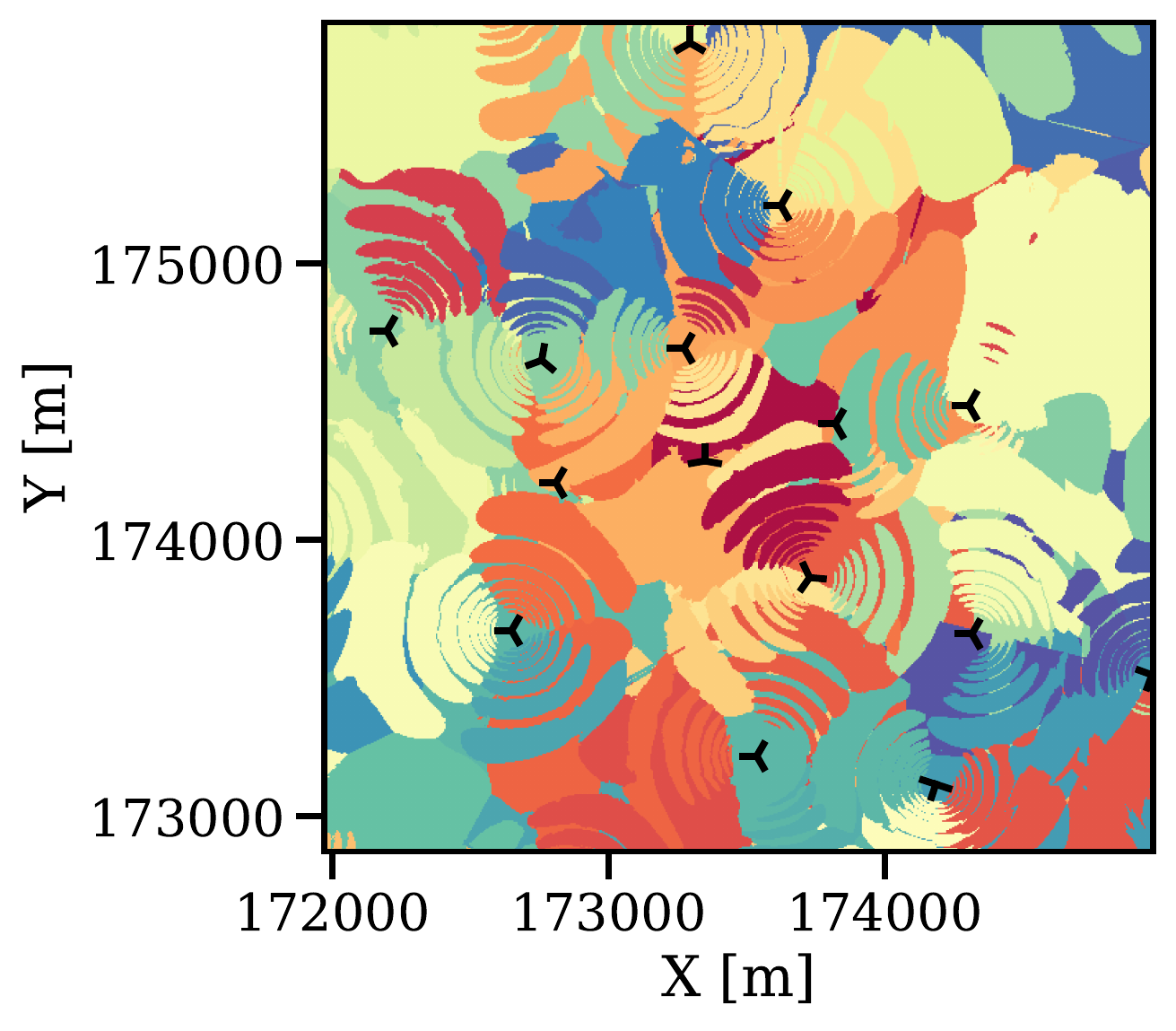}
        \caption{Operator 1, $120~m$}
    \end{subfigure}
    \vfill
	\begin{subfigure}[b]{0.49\linewidth}
    \centering
    \includegraphics[width=\textwidth]{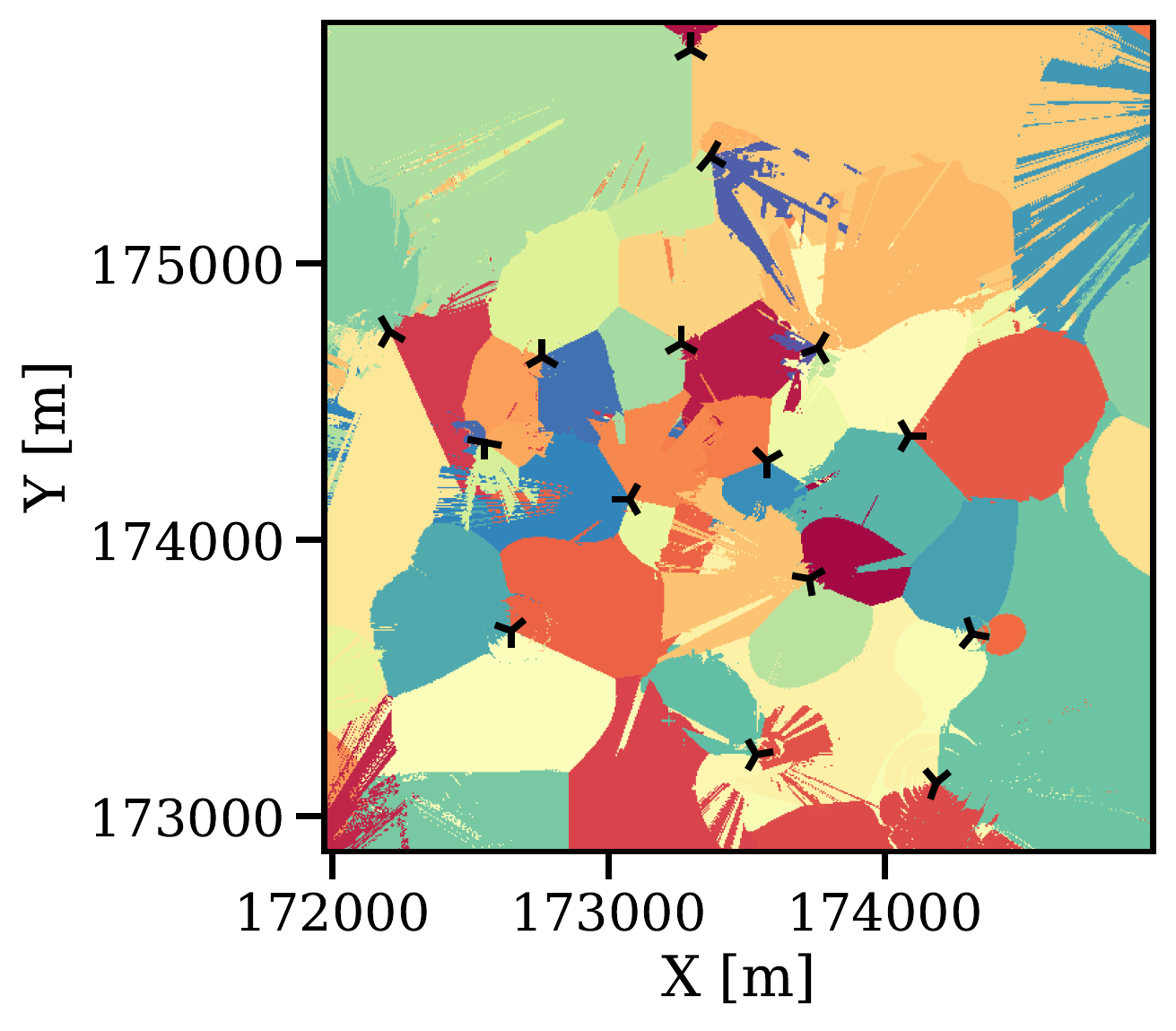}
        \caption{Operator 2, $20~m$}
    \end{subfigure}
    \hfill
	\begin{subfigure}[b]{0.49\linewidth}
    \centering
    \includegraphics[width=\textwidth]{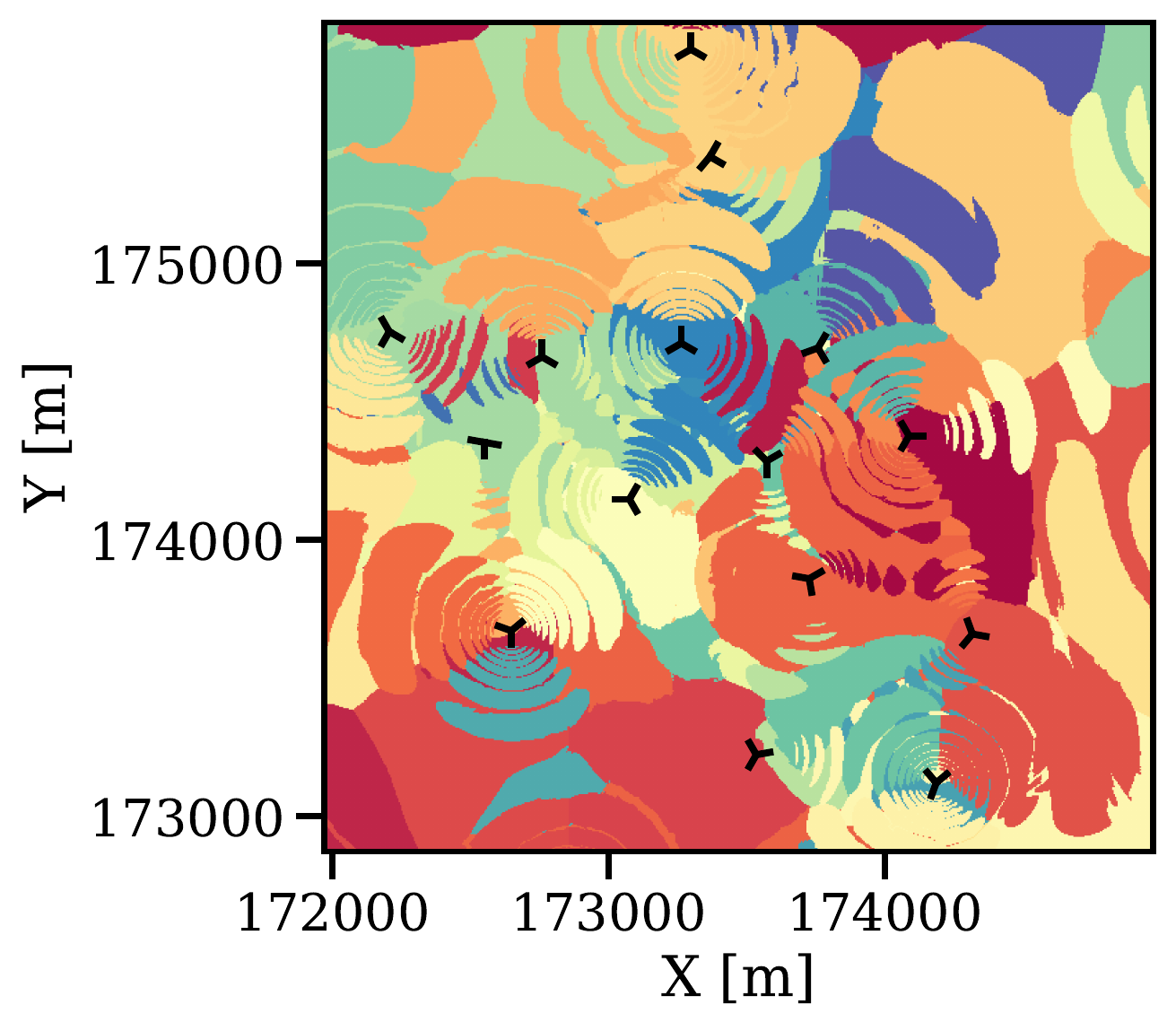}
        \caption{Operator 2, $120~m$}
    \end{subfigure}
	    \caption{Sector assignment map for two network operators at $20~m$ and $120~m$ height. Each color represents a different sector. The color of a pixel indicates which sector a user at this location is assigned to.}
     \label{fig:leuven_assmap}
\end{figure}

 The assignment map indicates which sector has the highest received power at each point. In Fig. \ref{fig:leuven_assmap} the assignment map of two operators at different altitudes is shown. Each color represents a sector, meaning that the color of a pixel determines which sector a user is assigned to at this location. One can observe that at $120~m$ height the assignment pattern is much more fragmented than the one at $20~m$ height. Due to the antenna sidelobes and nulls, UAVs at high altitudes will connect to sectors that are not necessarily the closest or they will alternate rapidly between multiple sectors. This will lead to higher handover frequency. These handovers result in throughput drops, where at these altitudes throughput is already a scarce resource. To build a reliable UAV communications network, reducing the number of handovers and Radio-Link-Failures is in the best of interest. 
Fig. \ref{fig:leuven_assmap} also shows the difference in assignment patterns between two different operators. We can utilize this diversity to create a multi-operator network where when a handover needs to happen or when a RLF occurs, the user can always fall back on the other operator's network.


\par
\begin{figure}[]
    \centering 
    
    \resizebox{0.825\linewidth}{!}{
    \begin{subfigure}{0.9\linewidth}
        \input{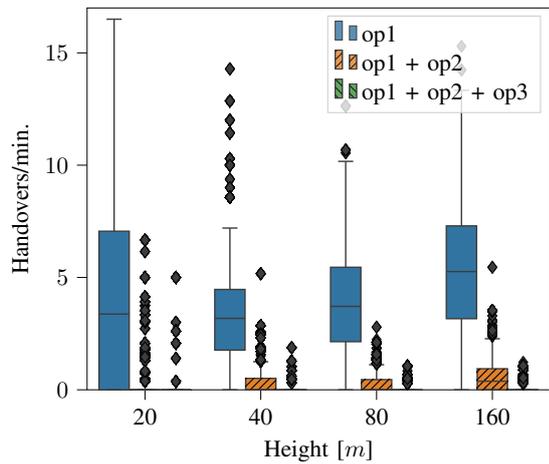}
        \caption{Number of handovers per minute}
        \label{fig:handovers_time}
    \end{subfigure} 
    }
    \resizebox{0.825\linewidth}{!}{
    \begin{subfigure}{0.9\linewidth}
        \input{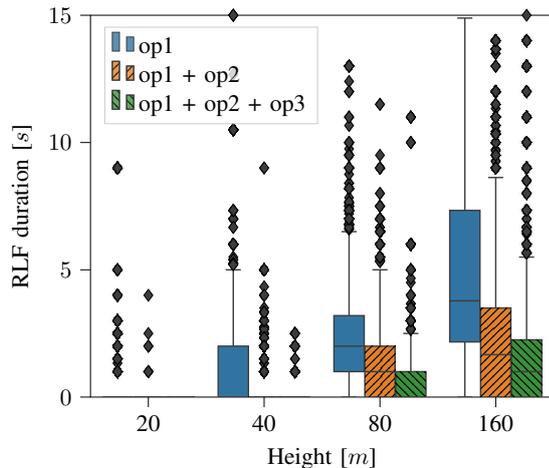}
        \caption{Radio Link Failure duration}
        \label{fig:rlf_dur}
    \end{subfigure} 
    }
    \caption{Impact of multi-operator diversity on the number of handovers and on the time spent disconnected due to RLF.}
	\label{fig:handovers_results}
\end{figure}

 The results can be neatly observed in the Fig. \ref{fig:handovers_results}. Fig. \ref{fig:handovers_time} shows the  handover frequency experienced by a drone flying at certain altitudes. When looking at a single operator system, we can see that a drone flying at an altitude of $160~m$ experiences a median of five handovers, this is a handover every 12 seconds. It is clear that this will cause significant problems for the command and control link. In contrast, the multi operator systems experience approximately zero handovers at the same altitude, due to the network diversity.
\par The same positive effect can be seen when monitoring the RLF duration. Fig. \ref{fig:rlf_dur} shows the duration a drone will not be connected to the network when it encounters a RLF. At an altitude of $160~m$ a drone using a single network will have a median disconnection time of $4~s$ when encountering a no-coverage zone. This is clearly not a desired situation. However, a drone connecting to multiple networks simultaneously can bring down this duration to a median of $1~s$. This shows the clear benefits for drones of connecting to multiple cellular networks at the same time.

\label{sec:handovers}



\section{Conclusion}
\par In this article an application of multi-operator network diversity has been explored to improve the network conditions for UAVs in deployed mobile networks. The results indicate that the increase in performance is indeed significant: the coverage can be improved by 20\% even in the worst case scenario of full network load, while the size of the outage zones is up to ten times smaller, particularly for high altitudes, than for the separate networks, meaning that the drone will have less chance to be unreachable for a long time due to bad network conditions.
The mobility characteristics can also be improved, the possible disconnection events can be reduced significantly, resulting in
a better overall latency for the combined connection. 

\bibliographystyle{IEEEtran}
\bibliography{new_references}
\end{document}